\documentclass[12pt]{article}

\usepackage{amssymb,amsmath,amsfonts,eurosym,geometry,ulem,graphicx,caption,color,setspace,sectsty,comment,footmisc,caption,natbib,pdflscape,subfigure,array,hyperref,mathtools,threeparttable}

\mathtoolsset{showonlyrefs=true}

\newcommand{\re}{\mathbb{R}}
\newcommand{\ex}{\mathbb{E}}
\newcommand{\pr}{\mathrm{Pr}}
\newcommand{\de}{\Delta}

\newcommand{\sumx}{\sum_{x'=1}^{J}}
\newcommand{\sumxm}{\sum_{x'=1}^{J-1}}
\newcommand{\sumi}{\sum_{i \in \mathcal{I}}}
\newcommand{\sumj}{\sum_{j=1}^{K}}
\newcommand{\sumw}{\sum_{w'=1}^{M}}
\newcommand{\exe}{\mathbb{E}_{\varepsilon}}
\newcommand{\sig}{\sigma_{t}^{*} \left( x, \epp \right)}
\newcommand{\xb}{\left( x \right)}
\newcommand{\wb}{\left( w \right)}
\newcommand{\xwb}{\left( x, w \right)}
\newcommand{\jb}{\left( J \right)}
\newcommand{\xbn}{\left( x' \right)}
\newcommand{\ep}{\varepsilon}
\newcommand{\epp}{\boldsymbol{\varepsilon}}
\newcommand{\sigg}{\boldsymbol{\sigma}}

\newcommand{\vt}{\mathbf{V}_{t+1}}
\newcommand{\vv}{\mathbf{V}}
\newcommand{\dd}{\mathbf{D}}
\newcommand{\pp}{\mathbf{P}}
\newcommand{\ii}{\mathbf{I}}
\newcommand{\ff}{\mathbf{F}}

\newcommand{\hf}{\mathbf{H}}
\newcommand{\ab}{\mathbf{A}}
\newcommand{\pf}{\mathbf{\Phi}}
\newcommand{\abt}{\tilde{\mathbf{A}}}
\newcommand{\Bt}{\tilde{\mathbf{B}}}

\newcommand{\ptil}{\tilde{\mathbf{P}}}
\newcommand{\bd}{\beta \delta}
\newcommand{\bdinv}{\left( \beta \delta \right)^{-1}}
\newcommand{\bfrac}{\frac{1-\beta}{\beta}}

\newcommand{\xcal}{\mathcal{X}}
\newcommand{\ical}{\mathcal{I}}
\newcommand{\wcal}{\mathcal{W}}
\newcommand{\el}{\mathcal{L}}
\newcommand{\theu}{\theta_{u}}
\newcommand{\thef}{\theta_{f}}
\newcommand{\hthef}{\hat{\theta}_{f}}
\newcommand{\ftil}{\tilde{\mathbf{F}}}
\newcommand{\finv}{\tilde{\mathbf{F}}^{-1}}

\newtheorem{thm}{Theorem}

\newtheorem{asm}{Assumption}

\newtheorem{example}{Example}

\newtheorem{rem}{Remark}

\newcolumntype{L}[1]{>{\raggedright\let\newline\\arraybackslash\hspace{0pt}}m{#1}}
\newcolumntype{C}[1]{>{\centering\let\newline\\arraybackslash\hspace{0pt}}m{#1}}
\newcolumntype{R}[1]{>{\raggedleft\let\newline\\arraybackslash\hspace{0pt}}m{#1}}

\begin{document}


\begin{titlepage} 
\title{Identifying Dynamic Discrete Choice Models with Hyperbolic Discounting \thanks{The author is grateful to Katsumi Shimotsu for his continuous guidance and support. The author is also grateful to Ruli Xiao for sharing the codes. The author would also like to thank Daiji Kawaguchi, Kosuke Uetake, Shintaro Yamaguchi, and participants in 2021 Asian Meeting of Econometric Society, 2021 Delhi Winter School of Econometric Society, and seminars at the University of Tokyo. The author gratefully acknowledges the support of JSPS KAKENHI Grant Number JP22J12532 and the Nakajima Foundation.}}
\author{Taiga Tsubota\thanks{University of Tokyo and Yale University.}}
\date{\today}
\maketitle
\begin{abstract}
\noindent We study identification of dynamic discrete choice models with hyperbolic discounting.
We show that the standard discount factor, present bias factor, and instantaneous utility functions for the sophisticated agent are point-identified from observed conditional choice probabilities and transition probabilities in a finite horizon model.
The main idea to achieve identification is to exploit variation in the observed conditional choice probabilities over time.
We present the estimation method and demonstrate a good performance of the estimator by simulation. \\
\vspace{0in}
\vspace{0in}\\

\bigskip
\end{abstract}
\setcounter{page}{1} 
\thispagestyle{empty} 
\end{titlepage} 
\pagebreak \newpage


\onehalfspacing

\section{Introduction} \label{sec: introduction}

Dynamic discrete choice (DDC) models have been widely used in applied microeconomics, such as industrial organization, health economics, labor economics, and political economy.
The DDC model is an econometric model that has a close link to economic theory, and hence, its main advantage is that the researcher can infer the mechanism of an agent's decision-making from data.
Using DDC models, the agent's preferences have been estimated in a broad range of social problems.
For example, \cite{keane1997career} analyze career decision-making, \cite{kennan2011effect} develop a DDC model on optimal migration, \cite{bayer2016dynamic} model neighborhood choice, and \cite{de2019subsidies} infer the agent's incentive to introduce renewable energy technologies.

Many DDC models build on dynamic optimization with exponential discounting, in which an agent discounts future streams of utility exponentially, so that the intertemporal marginal rate of substitution is constant over time.
Exponential discounting has been extensively used because it is simple to estimate the model and easy to interpret due to time-consistent preferences.

In real life, however, many cases of decision making do not follow such exponential discounting.
\cite{strotz1955myopia} and \cite{thaler1981some} provide an example: Most people tend to choose to wait for two apples in one year and a day, rather than one apple in one year, but some people choose to get one apple today rather than two apples tomorrow.
This evidence clearly contradicts exponential discounting.
\cite{thaler1981some} provides laboratory results that support this fact.
In addition, \citet*{frederick2002time} provide a review of the evidence for hyperbolic discounting, in which the agent has time-inconsistent preferences.

More importantly, in addition to experimental evidence, hyperbolic discounting is motivated in terms of policy implications.
This is because the estimation under different types of discounting can potentially yield quite different results.
For example, \cite{kennan2011effect} analyze migration patterns in the United States using a DDC model under exponential discounting.
They find that the variables in the moving costs, such as distance, home location, and population have significant effects on moving decisions.
However, if we re-estimate the model using hyperbolic discounting, the estimation result can potentially be altered.
This change may significantly impact policy decisions, such as subsidies for moving costs.
A more detailed explanation can be found in Examples \ref{ex:location} and \ref{ex:solarPV} in Section \ref{sec: model}. 

To the best of our knowledge, hyperbolic discounting is first modeled by \cite{strotz1955myopia}.
\cite{phelps1968second} and \cite{laibson1997golden} later remodel time-inconsistent preferences using a quasi-hyperbolic discounting model, in which the present preferences are represented by the following utility function:
\begin{align}
    U_{t} = u_{t} + \beta \delta u_{t+1} + \beta \delta^{2} u_{t+2} + \ldots + \beta \delta^{T-t} u_{T},
    \nonumber
\end{align}
where each $u_{\tau}$ represents instantaneous utility in period $\tau$, the constant $\delta$ is called the standard discount factor which discounts future streams of utility exponentially, and the constant $\beta$ is called the present bias factor which captures the agent's myopia. 
The product $\bd^{\tau-t}$ in period $\tau$ captures the degree of discounting future utility in the hyperbolic sense.
This quasi-hyperbolic discounting model includes exponential discounting as a special case.\footnote{Throughout, we refer to quasi-hyperbolic discounting simply as hyperbolic discounting for simplicity.}

Little is known about identification results on the primitives of DDC models with hyperbolic discounting.
Although some empirical papers (such as \cite{fang2009time} and \cite{paserman2008job}) try to estimate DDC models with hyperbolic discounting, they rely on parametric assumptions.
Such approaches are sensitive to the functional form and it has been unknown whether DDC models with hyperbolic discounting are identified without parametric assumptions.

This paper models the agent's decision making based on quasi-hyperbolic discounting, following \cite{fang2015estimating}.
We show that the present bias factor, the standard discount factor, and instantaneous utility functions are point-identified from the observed conditional choice probabilities (CCPs) and transition probabilities in the finite-horizon DDC model.

In the finite horizon framework, we achieve identification of the present bias factor and the standard discount factor by exploiting variation in the observed CCPs over time.
In the context of identification, because the observed CCPs and the state transition rules are observable from the data, we obtain a closed-form solution of the discount functions (the present bias factor and the standard discount factor) using these observables.
The strategy for identification is as follows.
First, we get a recursive expression of an agent's perceived long-run value function.
Second, in order to solve it, we take the logarithm of the CCP ratio and use the so-called Hotz-Miller inversion.
Third, we find pairs of actions that yield the same level of instantaneous utility and impose  stationary structures in instantaneous utility functions.
Then, all of the unknown variables are expressed in the form of functions of known variables, and hence, the model is identified.

We propose a maximum likelihood estimator to estimate the parameters of the model. 
Based on the identification results and the proposed estimation methods, we conduct Monte Carlo experiments and demonstrate a good performance of our estimator.

This paper contributes to the literature on the identification of the DDC models.
\cite{rust1994structural} shows the non-identifiability of primitives in the infinite horizon setting.
\cite{magnac2002identifying} show the degree of under-identification. 
In relation to this paper, they further claim that the standard discount factor is identified using exclusion restriction in the exponential discounting model.
\cite{kasahara2009nonparametric} and \cite{hu2012nonparametric} provide the nonparametric identification results, taking unobserved heterogeneity into account.
\cite{arcidiacono2020identifying} present the identification results focusing on the short panel setting like our models.
Using the exponential discounting model, \cite{abbring2020identifying} show set-identification results of the standard discount factor.
\citet*{an2021dynamic} ease the rational expectation assumption in DDC models and identify the agent's subjective beliefs on state transitions.
They study DDC models with exponential discounting and assume that discount functions are known.
On the other hand, we identify DDC models including discount functions with hyperbolic discounting, while maintaining the rational expectation assumption.
Last but not least, \cite{abbring2010identification} review the identification results of the DDC models.

This paper is closely related to \cite{fang2015estimating}, \citet*{abbring2018identifying}, and \citet*{wang2022identification} (WWX hereafter).
These papers show identification results in the hyperbolic discounting setting.
To the best of our knowledge, \cite{fang2015estimating} are the first who analyze the identification of the DDC model with hyperbolic discounting.
Although their identification results are refuted by \cite{abbring2020comment}, \cite{fang2015estimating} provide the basic framework that is useful for our model.
\citet*{abbring2018identifying} also consider the identification of DDC models with hyperbolic discounting, but they provide the set identification results.
In contrast, we provide the point identification results.

WWX also provide complementary results to ours: they present point identification results for both naive agents and sophisticated agents.\footnote{Agents are said to be \textit{sophisticated} if they are fully aware of their present bias and \textit{naive} if not.}
Notably, for sophisticated agents, they show that the discounting functions are identified if the final three periods are observed.
However, their results crucially depend on the existence of a terminating action.
On the other hand, our identification argument does not rely on terminating actions.
As a result, our framework can incorporate a wider range of DDC models, including decision processes without terminating actions.
For example, the dynamic analysis of the migration decision does not have a terminating action, as it is a life-long decision-making (see Example \ref{ex:location} in Section \ref{sec: model}). 
Although our identification result has an advantage in that it does not require terminating actions, our result requires a larger number of time periods in data.
In our framework, the number of time periods required in the data increases linearly as the number of the support of the state variable increases.
In contrast, WWX only require the final three periods to be observed.
Therefore, there is a trade-off between our result and WWX for identification in terms of the need for terminating actions and the required number of time periods.

The identification approach of our paper and WWX are similar: they both use a variation in CCPs over time and obtain the identification of discount functions by matrix inversion.
As a result, they both assume the full rank condition for the matrices related to the transition matrix and CCPs but in different forms.
Broadly speaking, this difference in ways of constructing matrices results in the difference in required assumptions.

The structure of this paper is as follows.
Section \ref{sec: model} presents the DDC model with hyperbolic discounting.
Section \ref{sec: identification} provides the main identification result.
Section \ref{sec:estimation} presents the estimation method, and section \ref{sec:simulation} shows the results of the simulation.
Section \ref{sec:conclusion} concludes.

\section{Model} \label{sec: model}

In this section, we consider a finite horizon dynamic discrete choice model where an agent has time-inconsistent preferences.
Our model is closely related to that of \cite{fang2015estimating}, but we assume that the time horizon is finite.
Our notation broadly follows that of \cite{fang2015estimating} and \citet*{an2021dynamic}.

In each discrete time period $t=1,2,\ldots,\bar{T}$ ($\bar{T} < \infty$), an agent chooses a discrete action from the set $\ical \coloneqq \left\{ 1,2,\ldots,K \right\}$ ($K<\infty$).
The agent's instantaneous utility depends on the action they choose and the set of state variables, where $x \in \xcal \coloneqq \left\{ 1,2,\ldots,J \right\}$ ($J<\infty$) is observed by the econometrician, and $\epp \coloneqq \left( \ep_{1}, \ldots, \ep_{K} \right) \in \re^{K}$ is a vector of choice-specific shocks that are not observed by the econometrician.
The agent can observe both the current state variables $x$ and $\epp$, and the agent chooses a current action so as to maximize the expected lifetime utility.
Given the current state variables $\left( x, \epp \right)$ and the agent's choice $i$, the state variables for the next period $(x', \epp')$ are realized by the exogenous transition function $f \left( x', \epp' \mid x, \epp, i \right)$.
We henceforth use a prime on a state variable when it denotes the state in the next period.
Regarding the state transition rule, we pose the following assumption.

\begin{asm} \label{asm: transition}
    The observed and unobserved state variables evolve independently, conditional on $x$ and $i$:
    \begin{align}
        f \left( x', \epp' \mid x, \epp, i \right)
        &= g \left( \epp' \mid x' \right) f \left( x' \mid x, i \right)
        \nonumber \\
        g \left( \epp' \mid x' \right)
        &= g \left( \epp' \right) \nonumber
    \end{align}
\end{asm}

In Assumption \ref{asm: transition}, we assume that the transition rule governs the evolution of the state variables in way of a Markov process given the agent's action.
We assume that the transition function is time-invariant, which is common in the literature. 

In the agent's dynamic optimization model, we implicitly assume that the agent has rational expectations or perfect expectations, that is, each agent realizes the true transition rule of both observed and unobserved state variables.
This assumption is standard in the literature of dynamic discrete choice models (e.g., \cite{magnac2002identifying}).
An important exception is the model of \citet*{an2021dynamic}; they consider the situation in which the agent has subjective beliefs about the law of motion of the observed state variables. 
Although their identification result provides a new insight into the literature, we stick to the assumption of the agent's perfect expectations.

We now make an assumption on the agent's instantaneous utility.

\begin{asm}
    The instantaneous utility is time-invariant and is given by
    \begin{align}
        u_{i}^{*} \left( x, \epp \right) 
        = u_{i} \xb + \ep_{i},
        \nonumber
    \end{align}
    for each $i \in \ical$.
    \label{asm: additive-separability}
\end{asm}

The additive separability assumption on the agent's instantaneous utility has been widely used in the literature (e.g., \cite{rust1987optimal}, \cite{hotz1993conditional}, and \cite{rust1994structural}).
The stationarity assumption in the finite horizon setting can be seen in the model of \citet*{an2021dynamic}.

There are two types of discount functions of the agent.
The first one is the standard discount factor $\delta \in (0,1)$, which denotes the agent's long-run discounting in every period.
This discounting function exponentially discounts future streams of utility, capturing the time-consistent part of the agent's preferences.
The second one is the present-bias factor $\beta \in (0,1)$, which denotes the agent's short-term impatience: it further discounts all the future utility from tomorrow on.
Following \cite{phelps1968second} and \cite{laibson1997golden}, we represent the agent's intertemporal preferences by
\begin{align}
    U_{t} \left( u_{t}, u_{t+1}, \ldots u_{\bar{T}} \right)
    = 
    u_{t} + \beta \sum_{\tau=t+1}^{\bar{T}} \delta^{\tau-t} u_{\tau},
    \nonumber
\end{align}
where $u_{\tau}$ is the agent's instantaneous utility in period $\tau$.
According to \cite{laibson1997golden}, this model is called quasi-hyperbolic discounting, because it approximates the hyperbolic discounting model.

Because the agent has time-inconsistent preferences, we consider the model as the game played by the current agent herself and the future selves.
Following \cite{fang2015estimating}, let $\sigma_{t}: \xcal \times \re^{K} \to \ical$ be a Markovian choice strategy and $\sigg_{t}^{+} \coloneqq \left\{ \sigma_{k} \right\}_{k=t}^{\bar{T}}$ the continuation strategy profile subsequent to period $t$.
Given $\sigg_{t}^{+}$, the agent's value function at period $t$ when the realized state variables are $x$ and $\epp$ is recursively written as
\begin{align}
    V_{t} \left( x, \epp; \sigg_{t}^{+} \right)
    = u_{\sigma_{t} \left(x, \epp \right)}^{*} \left( x, \ep_{\sigma_{t} \left( x, \epp \right)} \right) + \delta \ex \left[ V_{t+1} \left( x', \epp'; \sigg_{t+1}^{+} \right) \mid x, \sigma_{t} \left( x, \epp \right) \right].
    \label{eq: def-generic-value-function}
\end{align}
Following \cite{o1999doing}, \cite{o2001choice}, and \cite{fang2015estimating}, we define the perception-perfect strategy profile $\sigg^{*} \coloneqq \left\{ \sigma_{t}^{*} \right\}_{t=1}^{\bar{T}}$ such that
\begin{align}
    \sig
    = \arg \max_{i \in \ical} \left\{ u_{i}^{*} \left( x, \ep_{i} \right) + \beta \delta \ex \left[ V_{t+1} \left( x', \epp'; \sigg_{t+1}^{*+} \right) \mid x, i \right] \right\},
    \label{eq: def-perceived-strategy}
\end{align}
for all $t$, $x$, and $\epp$ (where $\sigg_{t}^{*+} \coloneqq \left\{ \sigma_{k}^{*} \right\}_{k=t}^{\bar{T}}$).
Using \eqref{eq: def-generic-value-function} and \eqref{eq: def-perceived-strategy}, we define the agent's perceived long-run value function as
\begin{align}
    V_{t} \xb
    = \exe \left[ V_{t} \left( x, \epp; \sigg_{t}^{*+} \right) \right],
    \label{eq: def-perceived-longrun-value-function}
\end{align}
for all $t$ and $x$.

With the perceived long-run value function, we now define two types of choice-specific value functions to analyze how the agent makes a decision.
Following \cite{fang2015estimating}, 
the perceived choice-specific long-run value function is defined as
\begin{align}
    V_{t,i} \xb
    = u_{i} \xb + \delta \sumx V_{t+1} \xbn f \left( x' \mid x, i \right),
    \label{eq: def-choice-specific-vti}
\end{align}
and the agent's current choice-specific value function is defined as
\begin{align}
    W_{t,i} \xb
    = u_{i} \xb + \bd \sumx V_{t+1} \xbn f \left( x' \mid x, i \right).
    \label{eq: def-choice-specific-wti}
\end{align}
The two types of choice-specific value functions differ in two ways. First, in $V_{t,i}\xb$, the value $s$ periods ahead from $t$ is discounted by $\delta^{s}$, while in $W_{t,i}\xb$, it is discounted by $\beta\delta^{s}$.
Second, the perceived choice-specific long-run value function, $V_{t,i}\xb$, captures the way of agent's discounting as if the agent ignores their own present-bias (i.e., $\beta=1$). That is, the agent is not aware of their own myopia in discounting. 
In contrast, the current choice-specific value function, $W_{t,i}\xb$, correctly captures the agent's discounting behavior including their own present bias.

We impose assumptions on the distribution of unobserved state variables and the conditional choice probability (CCP) $P_{t,i}\xb$, which is the probability that action $i$ is taken at period $t$ when the state variable $x$ is realized.

\begin{asm} \label{asm: logit-error}
    (a) The unobserved state variable $\epp$ is i.i.d. type 1 extreme value distributed.
    \\
    (b) The CCP $P_{t,i}\xb$ is known and determined by the current choice-specific value function $W_{t,i}\xb$.
\end{asm}

The extreme value distribution assumption in Assumption \ref{asm: logit-error} (a) is standard in the literature of dynamic discrete choice models (e.g., Assumption CLOGIT in \cite{aguirregabiria2010dynamic}).
Although there are some other important exceptions on the shock distribution such as the generalized extreme value distribution (e.g., \cite{arcidiacono2011conditional}) and the unknown distribution (e.g., \cite{norets2014semiparametric}), we maintain the assumption of type 1 extreme value distribution so as to get analytical convenience.
In Assumption \ref{asm: logit-error} (b), we consider the situation where $N\to\infty$ where $N$ is the number of agents in the data. This makes our identification arguments valid.

Under Assumption \ref{asm: logit-error} (a) and (b), the (observed) CCP $P_{t,i}\xb$ is given by
\begin{align}
    P_{t,i} \xb
    = \pr \left[ W_{t,i} \xb + \ep_{t,i} > \max_{j \in \ical \backslash \left\{ i \right\}} \left\{ W_{t,j} \xb + \ep_{t,j} \right\} \right]
    = \frac{\exp \left[ W_{t,i} \xb \right]}{\sumj \exp \left[ W_{t,j} \xb \right]}.
    \label{eq: def-observedCCP}
\end{align}

Before we present the identification result, we consider some empirical examples.

\begin{example} \label{ex:location}
    The choice of housing location has been widely studied. 
    \cite{kennan2011effect} analyze migration patterns in the United States, using a dynamic discrete choice model.
    In their model, each agent not only chooses whether or not to stay in their current location, but also chooses where to live if the agent decides to move.
    The location choices and state variables such as wages, moving costs, nonpecuniary amenity values, and a home premium, affect the agent's instantaneous payoff.
    \cite{kennan2011effect} set up a finite-horizon stationary model with exponential discounting.
    There are no terminating actions, so location choices are not once and for all.
    
    \cite{kennan2011effect} find that the variables in the moving costs, such as distance, home location, and population have significant effects on moving decisions.
    Their results are based on estimation under exponential discounting, assuming that the standard discount factor is known.
    The use of hyperbolic discounting and the estimation of discount functions have meaningful implications for policy analysis.
    This is because the estimation results would be different if we estimate the model under hyperbolic discounting with unknown discount factors. 
    Consequently, the new model and its estimation results could potentially alter policy decisions on subsidies for moving costs, for example.
\end{example}

\begin{example} \label{ex:solarPV}
    \cite{de2019subsidies} analyze a subsidy program designed to promote the adoption of solar photovoltaic (PV) systems.
    The subsidies are provided for future electricity production, not for upfront investment.
    They develop a DDC model and identify the standard discount factor. 
    In their model, each household makes a discrete choice in each period: which PV alternative to choose or not to adopt. 
    This choice influences both instantaneous utility and future states. 
    State variables include the upfront investment cost of the PV system, electricity cost savings by adoption, and subsidy levels. 
    In this example, adopting one brand of PV systems is a terminating action.

    \cite{de2019subsidies} find that households significantly discount future benefits under exponential discounting, which indicates the potential benefit of introducing hyperbolic discounting.
    This issue is of crucial importance in policy analysis.
    Under exponential discounting, they find that an upfront investment subsidy program would have been more cost-effective than subsidies for future electricity production.
    The results of the policy analysis would change significantly if hyperbolic discounting is introduced, thereby affecting the design of the subsidy program by policy makers.
\end{example}

\section{Identification} \label{sec: identification}

In this section, we show that, in finite horizon framework, discount functions $\beta$ and $\delta$ are identified using time variation in observed CCPs.
Our identification argument is close to that of \citet*{an2021dynamic}.
They study DDC models with exponential discounting and relax the rational expectation assumption to identify the agent's subjective beliefs on state transitions, while assuming that discount functions are known.
On the other hand, we identify DDC models including discount functions with hyperbolic discounting, while maintaining the rational expectation assumption.

To provide a brief overview of our identification strategy, note that there are three types of value functions, $V_{t}\xb$, $V_{t,i}\xb$, and $W_{t,i}\xb$, as introduced in Section \ref{sec: model}. We exploit the relationships between these value functions and their connection to the observed CCPs, $P_{t,i}\xb$. By taking time differences and appropriately taking differences from a reference state, we can separate the discounting functions from the observables. Under the assumptions of full rank matrices and the existence of pairs of actions or states that yield the same utility level, both $\beta$ and $\delta$ are individually derived in closed form.

Let $T$ be the last period of the data ($T \le \bar{T}$).
First, note that, by \eqref{eq: def-perceived-longrun-value-function} and \eqref{eq: def-choice-specific-vti}, we obtain 
\begin{align}
    V_{t} \xb
    = \exe \left[ V_{t, \sig} \xb + \ep_{t, \sig} \right],
    \label{eq: rel-valuefunction-choicespecific}
\end{align}
which will be useful for identification.
By combining equations \eqref{eq: def-choice-specific-vti} and \eqref{eq: def-choice-specific-wti}, we have
\begin{align}
    V_{t,i} \xb - W_{t,i} \xb
    &= \left( 1 - \beta \right) \delta \sumx V_{t+1} \xbn f \left( x' \mid x, i \right).
    \label{eq: difference-vti-wti}
\end{align}
Exploiting this relationship as well as equation \eqref{eq: rel-valuefunction-choicespecific} and Assumption \ref{asm: logit-error}, we have
\begin{align}
    V_{t} \xb
    &= \exe \left[ V_{t, \sig} \xb + \ep_{t, \sig} \right]
    \nonumber \\
    &= \exe \left[ W_{t, \sig} \xb + \ep_{t, \sig} + \left( 1-\beta \right) \delta \sumx V_{t+1} \xbn f \left( x' \mid x, \sig \right) \right]
    \nonumber \\
    &= \exe \max_{i} \left\{ W_{t,i} \xb + \ep_{t,i} \right\} + \left( 1-\beta \right) \delta \exe \left[ \sumx V_{t+1} \xbn f \left( x' \mid x, \sig \right) \right]
    \nonumber \\
    &= \log \left\{ \sumi \exp \left[ W_{t,i} \xb \right] \right\} + \left( 1-\beta \right) \delta \sumj \left[ \pr \left\{ \sig = j \right\} \sumx V_{t+1} \xbn f \left( x' \mid x, j \right) \right]
    \nonumber \\
    &= -\log \left[ P_{t,K} \xb \right] + W_{t,K} \xb + \left( 1-\beta \right) \delta \sumj P_{t,j} \xb \left[ \sumx V_{t+1} \xbn f \left( x' \mid x, j \right) \right].
    \label{eq: key-before-scalar-valuefunction}
\end{align}
By equation \eqref{eq: def-choice-specific-wti}, this equation can be rewritten as
\begin{align}
    V_{t} \xb
    = 
    &-\log \left[ P_{t,K} \xb \right] + u_{K} \xb + \beta \delta \sumx V_{t+1} \xbn f \left( x' \mid x, K \right)
    \nonumber \\
    &+ \left( 1-\beta \right) \delta \sumj P_{t,j} \xb \left[ \sumx V_{t+1} \xbn f \left( x' \mid x, j \right) \right].
    \label{eq: key-scalar-valuefunction}
\end{align}
If we take a difference against the reference state $J$, we obtain
\begin{align}
    V_{t} \xb - V_{t} \jb
    = 
    &- \left[ \log \left( P_{t,K} \xb \right) - \log \left( P_{t,K} \jb \right) \right]
    + \left[ u_{K} \xb - u_{K} \jb \right]
    \nonumber \\
    &+ \bd \left[ \sumx V_{t+1} \xbn f \left( x' \mid x, K \right) - \sumx V_{t+1} \xbn f \left( x' \mid J, K \right) \right]
    \nonumber \\
    &+ \left( 1 - \beta \right) \delta \sumj P_{t,j} \xb \left[ \sumx V_{t+1} \xbn f \left( x' \mid x, j \right) \right]
    \nonumber \\
    &- \left( 1 - \beta \right) \delta \sumj P_{t,j} \jb \left[ \sumx V_{t+1} \xbn f \left( x' \mid J, j \right) \right].
    \label{eq: key-scalar-difference-valuefunction}
\end{align}

In order to facilitate subsequent arguments, we introduce some notations.
Let 
\begin{align}
    \vv_{t} 
    = 
    \left[ 
    \begin{array}{c}
        V_{t} \left( x=1 \right) - V_{t} \jb
        \\ \vdots
        \\ V_{t} \left( x = J-1 \right) - V_{t} \jb
    \end{array}
    \right]
\end{align}
be a $(J-1) \times 1$ vector and define $\log \left( \pp_{t,K} \right)$ and $\mathbf{u}_{K}$ similarly.
Let 
\begin{align}
    \tilde{\ff}_{K} 
    = 
    \left[
    \begin{array}{c}
        \ff_{K} \left( x=1 \right) - \ff_{K} \jb
        \\ \vdots
        \\ \ff_{K} \left( x=J-1 \right) - \ff_{K} \jb
    \end{array}
    \right]
\end{align}
be a $(J-1) \times (J-1)$ matrix, where $\ff_{i} \xb = \left[ f \left( x'=1 \mid x,i \right), \ldots, f \left( x'=J-1 \mid x,i \right) \right]$ is a $1 \times (J-1)$ vector.
Let 
\begin{align}
    \ff 
    =
    \left[
    \begin{array}{c}
        \ff_{1}
        \\ \vdots
        \\ \ff_{K}
    \end{array}
    \right]
    \quad
    \text{and}
    \quad
    \ff_{J} 
    =
    \left[
    \begin{array}{c}
        \ff_{1,J} 
        \\ \vdots
        \\ \ff_{K,J} 
    \end{array}
    \right]
\end{align}
be $(J-1)K \times (J-1)$ matrices, where
\begin{align}
    \ff_{i} 
    =
    \left[
    \begin{array}{c}
        \ff_{i} \left( x=1 \right)
        \\ \vdots
        \\ \ff_{i} \left( x=J-1 \right)
    \end{array}
    \right]
    \quad
    \text{and}
    \quad
    \ff_{i,J} 
    =
    \left[
    \begin{array}{c}
        \ff_{i} \jb
        \\ \vdots
        \\ \ff_{i} \jb
    \end{array}
    \right]
\end{align}
are $(J-1) \times (J-1)$ matrices.
We also let
\begin{align}
    \pp_{t}
    =
    \left[
    \begin{array}{ccc}
        \pp_{t,1} & \cdots & \pp_{t,K}
    \end{array}
    \right]
    \nonumber
\end{align}
be a $(J-1) \times (J-1)K$ matrix where
\begin{align}
    \pp_{t,i}
    =
    \left[
    \begin{array}{cccc}
        P_{t,i} \left( x=1 \right) & 0 & \cdots & 0 \\
        0 & P_{t,i} \left( x=2 \right) & \cdots & 0 \\
        \vdots & \vdots & \ddots & \vdots \\
        0 & 0 & \cdots & P_{t,i} \left( x=J-1 \right)
    \end{array}
    \right],
    \nonumber
\end{align}
and let
\begin{align}
    \pp_{t,J}
    =
    \left[
    \begin{array}{ccc}
        \pp_{t,1,J} \cdots \pp_{t,K,J}
    \end{array}
    \right]
    \nonumber
\end{align}
be a $(J-1) \times (J-1)K$ matrix where
\begin{align}
    \pp_{t,i,J}
    =
    \left[
    \begin{array}{cccc}
        P_{t,i} \jb & 0 & \ldots & 0 \\
        0 & P_{t,i} \jb & \ldots & 0 \\
        \vdots & \vdots & \ddots & \vdots \\
        0 & 0 & \ldots & P_{t,i} \jb
    \end{array}
    \right].
    \nonumber
\end{align}
Using the above matrices, the right-hand side of equation \eqref{eq: key-scalar-difference-valuefunction} can be rewritten as 
\begin{align}
    \vv_{t}
    =
    &- \log \left( \pp_{t,K} \right)
    + \mathbf{u}_{K}
    + \bd \tilde{\ff}_{K} \vt
    + \left( 1-\beta \right) \delta \pp_{t} \ff \vt
    - \left( 1-\beta \right) \delta \pp_{t,J} \ff_{J} \vt.
    \label{eq: Vt-key}
\end{align}

Now we consider the CCP ratio to acquire another form of the first difference of the perceived long-run value function vector.
For actions $k,\ell \in \ical$ and states $x_{1}, x_{2} \in \xcal$, we define
\begin{align}
    D_{t,k,\ell} \left( x_{1}, x_{2} \right)
    &\coloneqq \log \left( \frac{P_{t,k} \left( x_{1} \right)}{P_{t,\ell} \left( x_{2} \right)} \right)
    \nonumber \\
    &= W_{t,k} \left( x_{1} \right) - W_{t,\ell} \left( x_{2} \right)
    \nonumber \\
    &= \left[ u_{k} \left( x_{1} \right) - u_{\ell} \left( x_{2} \right) \right] + \bd \sumx V_{t+1} \xbn \left[ f \left( x' \mid x,k \right) - f \left( x' \mid x,\ell \right) \right]
    \nonumber \\
    &= \left[ u_{k} \left( x_{1} \right) - u_{\ell} \left( x_{2} \right) \right] + \bd \left[ \ff_{k} \left( x_{1} \right) - \ff_{\ell} \left( x_{2} \right) \right] \vt,
    \label{eq: CCPratio}
\end{align}
where the second equality comes from the result of Proposition 1 in \cite{hotz1993conditional} and the fourth equality follows from the derivation in Appendix.

In equation \eqref{eq: CCPratio}, suppose that we have $(J-1)$ pairs of actions $\left( k,\ell \right)$ and states $\left( x_{1}, x_{2} \right)$ and let each pair be $\left( k^{j}, \ell^{j} \right)$ and $\left( x_{1}^{j}, x_{2}^{j} \right)$ for $j = 1,\ldots,J-1$.
Then we define the following $\left( J-1 \right) \times \left( J-1 \right)$ matrix:
\begin{align}
    \ftil
    = \left[
        \begin{array}{c}
            \ff_{k^{1}} \left( x_{1}^{1} \right) - \ff_{\ell^{1}} \left( x_{2}^{1} \right) \\
            \vdots \\
            \ff_{k^{J-1}} \left( x_{1}^{J-1} \right) - \ff_{\ell^{J-1}} \left( x_{2}^{J-1} \right)
        \end{array}
    \right]
\end{align}

In order to get a closed form of the ex-ante value function, we impose the following assumption:
\begin{asm}
    (a) There exist at least $J-1$ action pairs $\left( k^{j}, \ell^{j} \right)$ and state pairs $\left( x_{1}^{j}, x_{2}^{j} \right)$ such that: 
    \begin{align}
        u_{k^{j}} \left( x_{1}^{j} \right) = u_{\ell^{j}} \left( x_{2}^{j} \right),
    \end{align}
    with either (i) $k^{j} \ne \ell^{j}$, (ii) $x_{1}^{j} \ne x_{2}^{j}$, or (iii) both, for $j = 1, \ldots, J-1$.
    \\
    (b) the matrix $\ftil$ is full column rank.
    \label{asm: ukul}
\end{asm}

In part (a) of Assumption \ref{asm: ukul}, the researcher must find $J-1$ pairs of different actions or different states that give the exact same level of instantaneous utility.
There are two types of justification for this assumption.
First, this assumption includes normalization, which is commonly assumed in the DDC literature.
For example, normalization ($u_{K} \xb = 0$ for all $x$) is imposed in \cite{fang2015estimating} and \cite{abbring2020identifying} for DDC models; \cite{bajari2015identification} and \cite{aguirregabiria2020identification} for dynamic games.
\citet*{an2021dynamic} also assume it for the infinite-horizon two-state DDC model.
The reason why Assumption \ref{asm: ukul} (a) is more general than normalization is as follows.
Our assumption only requires that there exist $J-1$ pairs of states and actions that give the same utility level, and such pairs can be even across states and across actions.
On the other hand, normalization requires that all such pairs must be on the reference action $K$.
Second, this assumption is easily imposed in the real empirical analysis, because it is directly imposed on the instantaneous utility.
\cite{abbring2020identifying} also impose the similar assumption to ours, in the context of exponential discounting.
As they mention in their paper, this type of assumption has superiority to that in \cite{magnac2002identifying}.
This is because \cite{magnac2002identifying} impose an assumption on the current value function, which is often difficult to get an economic intuition.

The intuition for part (b) of Assumption \ref{asm: ukul} is as follows: when $J=2$, this condition reduces to $f \left( x'=1 \mid x_{1}, k \right) \ne f \left( x'=1 \mid x_{2}, \ell \right)$ for some $k, \ell \in \ical$ and some $x_{1}, x_{2} \in \xcal$.
Furthermore, when $x_{1} = x_{2}$, this is simplified to $f \left( x'=1 \mid x=1, k \right) \ne f \left( x'=1 \mid x=1, \ell \right)$.
This means that, starting from the same state, the transition probability to the identical state must be different for the different action.
When $J \ge 3$, this assumption requires that every row vector of $\ftil$ be linearly independent, which is easily testable.
In the literature of identification of DDC models, \citet*{an2021dynamic} introduce a similar assumption for the subjective beliefs for the state transition.

Assumption \ref{asm: ukul} makes it possible to express $\vt$ in a closed form. 
When the researcher can find more than $J-1$ action and state pairs, Assumption \ref{asm: ukul} (b) ensures that $\ftil$ has an inverse matrix and the subsequent argument is also valid.
Henceforth, we focus on the situation where we have exactly $J-1$ pairs in Assumption \ref{asm: ukul} (a) and $\ftil$ has an inverse matrix $\ftil^{-1}$.
Assumption \ref{asm: ukul} and equation \eqref{eq: CCPratio} enables us to obtain the following equations:
\begin{align}
    \vv_{t+1} 
    &= \bdinv \finv \dd_{t}
    \label{eq: key2-V-afterinv} \\ 
    \de \vt
    &= \bdinv \finv \de \dd_{t},
    \label{eq: key2-deltaV-afterinv}
\end{align}
where $\dd_{t} = \left[ D_{t,k^{1},\ell^{1}} \left( x_{1}^{1}, x_{2}^{1} \right), \ldots, D_{t,k^{J-1},\ell^{J-1}} \left( x_{1}^{J-1}, x_{2}^{J-1} \right) \right]'$ is a $(J-1) \times 1$ vector.
Taking the first difference of equation \eqref{eq: Vt-key}, applying equations \eqref{eq: key2-V-afterinv} and \eqref{eq: key2-deltaV-afterinv}, and noting that the instantaneous utility is time-invariant by Assumption \ref{asm: additive-separability}, we have that, for $t=3,\ldots,T$, 
\begin{align}
    &\ 
    \bdinv \finv \de \dd_{t-1}
    \nonumber \\
    &= 
    - \de \log \left( \pp_{t,K} \right) 
    + \tilde{\ff}_{K} \finv \de \dd_{t}
    + \bfrac \left( \pp_{t} \ff - \pp_{t,J} \ff_{J} \right) \finv \dd_{t}
    \nonumber \\
    &\ - \bfrac \left( \pp_{t-1} \ff - \pp_{t-1,J} \ff_{J} \right) \finv \dd_{t-1}.
    \label{eq: key3-longform}
\end{align}
The right-hand side of equation \eqref{eq: key3-longform} can be equivalently expressed as 
\begin{align}
    \left[
    \begin{array}{ccc}
        \ii_{J-1} & \bfrac \ii_{J-1}  & - \bdinv \ii_{J-1}
    \end{array}
    \right]
    \ab
    &=
    \de \log \left( \pp_{K} \right),
    \label{eq: key3-matrixform}
\end{align}
where
\begin{align}
    \ab
    &=
    \left[ 
    \begin{array}{ccc}
        \tilde{\ff}_{K} \finv \de \dd_{3} & \cdots & \tilde{\ff}_{K} \finv \de \dd_{T} \\
        \pf_{3} & \cdots & \pf_{T} \\
        \finv \de \dd_{2} & \cdots & \finv \de \dd_{T-1}
    \end{array}
    \right]
    \nonumber
\end{align}
is a $3(J-1) \times (T-2)$ matrix where 
\begin{align}
    \pf_{t} = \left( \pp_{t} \ff - \pp_{t,J} \ff_{J}  \right) \finv \dd_{t} - \left( \pp_{t-1} \ff - \pp_{t-1,J} \ff_{J} \right) \finv \dd_{t-1}
    \nonumber
\end{align}
and
$\de \log \left( \pp_{K} \right) = \left[ \de \log \left( \pp_{3,K} \right), \ldots, \de \log \left( \pp_{T,K} \right) \right]$ is a $(J-1) \times (T-2)$ matrix.

For identification, we make the following assumption:
\begin{asm}
    (a) The number of time periods in the data, $T$, satisfies $T \ge 3J-1$.
    \\
    (b) The matrix $\ab$ is full row rank.
    \label{asm: rightinverse}
\end{asm}

Assumption \ref{asm: rightinverse} requires the matrix $\ab$ to have the right inverse matrix. 
As a result, both $\bfrac \ii_{J-1}$ and $-\bdinv \ii_{J-1}$ in equation \eqref{eq: key3-matrixform} have a closed-form solution and thus identification is made possible. 
Since both the state transition functions and the observed CCPs are known, Assumption \ref{asm: rightinverse} is empirically testable. Remark \ref{rem: matrix-A} below (at the end of this section) demonstrates how to interpret the matrix $\ab$ and the full row rank condition \ref{asm: rightinverse} (b).

If we impose Assumptions \ref{asm: transition} to \ref{asm: rightinverse}, we have the following theorem on identification:
\begin{thm}
    Suppose that Assumptions \ref{asm: transition} to \ref{asm: rightinverse} hold.
    Then the discount functions $\beta$ and $\delta$ are identified for $t=1,2,\ldots,T$ with $T \ge 3J-1$.
    Furthermore, if the utility level of one action for one state is known, then the instantaneous utility functions $u_{i} \xb$ for all $i \in \ical$ and all $x \in \xcal$ are identified.
    \label{thm: onestate-identification-betadelta}
\end{thm}
The proof for Theorem \ref{thm: onestate-identification-betadelta} is in Appendix.
If we impose some degree of stationarity in the finite horizon framework, the present-bias factor and the standard discount factor are identified as a function of the observed CCPs and the transition density. 
Also, if we set the continuation value at the terminal value to be zero, due to the stationary structure of the instantaneous utility, each instantaneous utility function is identified.
Note that our identification strategy is different from the one proposed by \cite{fang2015estimating} or the one discussed by \cite{abbring2020comment}.



\begin{rem} \label{rem: matrix-A}
    We illustrate the interpretation of matrix $\ab$ and Assumption \ref{asm: rightinverse} using a simplified example with $K=2$ (binary action), $J=2$ (two states), and $T=5$ (five periods, which satisfies Assumption \ref{asm: rightinverse} (a)). 
    Let $\pi_{t}\left(x,x^{\prime}\right)$ be the transition probability from state $x$ at period $t$ to state $x^{\prime}$ at period $t+1$.
    Focusing on the first column of $\ab$, some algebra yields:
        \begin{align}
            \ab = \beta\delta
                \left[ 
                    \begin{array}{ccc}
                        \left[f(1|1,2)-f(1|2,2)\right]\left[V_{4}(1)-V_{4}(2)-V_{3}(1)+V_{3}(2)\right] & * & * \\
                        \left[\pi_{3}\left(1,1\right) - \pi_{3}\left(2,2\right)\right]\left[V_{4}(1)-V_{4}(2)\right] - \left[\pi_{2}\left(1,1\right) - \pi_{2}\left(2,2\right)\right] \left[ V_{3}(1)-V_{3}(2) \right] & * & * \\
                        V_{3}(1)-V_{3}(2)-V_{2}(1)+V_{2}(2) & * & *
                    \end{array}
                \right],
                \nonumber
        \end{align}
    where $*$ represents non-zero entries, defined analogously for $t=4,5$.
    Assume that $f(1|1,2)-f(1|2,2) \ne 0$ (corresponding to Assumption \ref{asm: ukul}(b) in this context). Then the full row rank condition \ref{asm: rightinverse} (b) requires sufficient variation in the value functions $V_{t}(x)$ over time.
    To see the opposite case, consider the situation where there is no variation in value functions.
    If $V_{t}(x) = V_{t+1}(x)\ (t=2,3,4)$ for example, then the last row of $\ab$ would be a zero row vector. Consequently, matrix $\ab$ would not have an inverse, and thus our identification argument would fail.
    Therefore, Assumption \ref{asm: rightinverse} effectively rules out cases where value functions remain constant over time, ensuring the necessary variation for identification.
\end{rem}

\section{Estimation} \label{sec:estimation}

We have presented the identification results so far and we can directly apply such identification results to estimation.
However, as can be seen in Theorem \ref{thm: onestate-identification-betadelta} and its proof, we have to use the right inverse of a matrix, and such a method can be difficult to use when a matrix is near singular.
Instead, we present a maximum likelihood estimator to estimate the standard discount factor and the present bias factor.

Let the data have $N$ agents and $T$ periods. 
Let the data be $\left\{ a_{nt}, x_{nt} \right\}$, where $a_{nt}$ is an action that the agent $n$ chooses at period $t$ and $x_{nt}$ is the corresponding state.

The likelihood function of the data is 
\begin{align}
    &\ 
    \el \left( x_{2}, \ldots, x_{T}, a_{1}, \ldots, a_{T} \mid x_{1}; \theu, \thef, \beta, \delta \right)
    \nonumber \\
    &= \prod_{n=1}^{N} \prod_{t=2}^{T} P_{t} \left( a_{nt} \mid x_{nt} ; \theu, \thef, \beta, \delta \right) f \left( x_{nt} \mid x_{n,t-1}, a_{n,t-1} ; \thef \right) P_{1} \left( a_{n1} \mid x_{n1} ; \theu, \thef, \beta, \delta \right),
    \nonumber
\end{align}
where $\theu$ is the parameter in the instantaneous utility, $\thef$ is the parameter in the state transition, and $P_{t} \left( a_{nt} \mid x_{nt} ; \theu, \thef, \beta, \delta \right)$ is the CCP of agent $n$ at period $t$.
The log-likelihood function of the data is 
\begin{align}
    &\ 
    \log \el \left( x_{2}, \ldots, x_{T}, a_{1}, \ldots, a_{T} \mid x_{1}; \theu, \thef, \beta, \delta \right)
    \nonumber \\
    &= \sum_{n=1}^{N} \sum_{t=1}^{T} \log P_{t} \left( a_{nt} \mid x_{nt} ; \theu, \thef, \beta, \delta \right) 
    + \sum_{n=1}^{N} \sum_{t=2}^{T} \log f \left( x_{nt} \mid x_{n,t-1}, a_{n,t-1} ; \thef \right),
    \nonumber
\end{align}
so that we can estimate $\thef$ separately from the other unknown parameters $\theu$, $\beta$, and $\delta$.

First, we can obtain an estimator $\hthef$ of $\thef$ by maximizing the following objective function:
\begin{align}
    \sum_{n=1}^{N} \sum_{t=2}^{T} \log f \left( x_{nt} \mid x_{n,t-1}, a_{n,t-1} ; \thef \right).
    \nonumber
\end{align}

Second, after obtaining $\hthef$, we estimate agents' CCPs by backward induction.
At the terminal period $T$, let the continuation value be zero so that the choice-specific value equals the instantaneous utility.
Then, we can date back and obtain the CCPs for all periods.
Using the estimated CCPs and $\hthef$, we can estimate the unknown parameters $\theu$, $\beta$, and $\delta$ by maximizing the following function:
\begin{align}
    \sum_{n=1}^{N} \sum_{t=1}^{T} \log P_{t} \left( a_{nt} \mid x_{nt} ; \theu, \hthef, \beta, \delta \right).
    \nonumber
\end{align}

\section{Simulation} \label{sec:simulation}

We present the results of Monte Carlo experiments based on the identification results and the estimation methods that we have described.
We consider a model with one state variable that can take five possible values ($\xcal = \left\{ 0,1,2,3,4 \right\}$) and with two choices ($\ical = \left\{ 1,2 \right\}$) in the finite horizon model ($T=16$, which satisfies Assumption \ref{asm: rightinverse} (a)).
We parameterize the instantaneous utility functions as $u_{1} \xb = \alpha_{0} + \alpha_{1} x$ for all $x \in \xcal$ and assume that $u_{2} \xb = 0$ for all $x \in \xcal$, maintaining Assumption \ref{asm: ukul} (a).
We set $\alpha_{0} = 0.5$ and $\alpha_{1} = -0.2$.
Also, each choice-specific shock $\ep_{i}$ is independently drawn from type 1 extreme value distribution with mean zero.
Moreover, each state transition matrix is set so that each element is randomly generated from the uniform distribution over $\left[ 0,1 \right]$ and each row sums to one.

We generate data by the following procedure.
First, we set the instantaneous utility functions and the transition probabilities as described above, and the discount functions as below.
Second, by backward induction, we calculate the ex-ante value function $V_{t+1} \xb$, the choice-specific value $W_{t,i} \xb$, and the conditional choice probability $P_{t,i} \xb$ for each $x,i$, and $t=1,\ldots,T$.
Third, we simulate each agent's action using the transition probabilities and conditional choice probabilities.
Given the generated data, we estimate $\alpha_{0}$, $\alpha_{1}$, $\delta$, and $\beta$ by jointly maximizing the likelihood function.

As for the discount functions, we set two sets of the values: 
\begin{align}
    (1) &\ \left( \delta, \beta \right) = \left( 0.9, 0.85 \right),
    \nonumber
    \\
    (2) &\ \left( \delta, \beta \right) = \left( 0.75, 0.7 \right).
    \nonumber
\end{align}

    \begin{table}
        \centering
        \begin{tabular}{cccccc}
            \\ \hline
             & True & $N=2000$ & $N=4000$ & $N=8000$ & $N=10000$ \\ \hline
            $\alpha_{0}$ & 0.5 & 0.494 & 0.495 & 0.496 & 0.496 \\
            & & (0.026) & (0.019) & (0.014) & (0.012) \\
            $\alpha_{1}$ & -0.2 & -0.199 & -0.199 & -0.199 & -0.199 \\
            & & (0.009) & (0.007) & (0.005) & (0.004) \\ \hline
            $\delta$ & 0.9 & 0.819 & 0.861 & 0.886 & 0.890 \\
            & & (0.288) & (0.218) & (0.163) & (0.152) \\
            $\beta$ & 0.85 & 0.795 & 0.819 & 0.827 & 0.830 \\
            & & (0.302) & (0.252) & (0.209) & (0.197) \\ \hline
        \end{tabular}
        \caption{
        Simulation results.
        The means and standard errors are computed from 10000 random simulation samples.
        }
        \label{tab:simulation1}
    \end{table}

    \begin{table}
        \centering
        \begin{tabular}{cccccc}
            \\ \hline
             & True & $N=2000$ & $N=4000$ & $N=8000$ & $N=10000$ \\ \hline
            $\alpha_{0}$ & 0.5 & 0.498 & 0.497 & 0.497 & 0.497 \\
            & & (0.026) & (0.019) & (0.014) & (0.013) \\
            $\alpha_{1}$ & -0.2 & -0.200 & -0.199 & -0.199 & -0.199 \\
            & & (0.009) & (0.007) & (0.005) & (0.004) \\ \hline
            $\delta$ & 0.75 & 0.677 & 0.715 & 0.753 & 0.759 \\
            & & (0.369) & (0.327) & (0.277) & (0.265) \\
            $\beta$ & 0.7 & 0.670 & 0.689 & 0.695 & 0.701 \\
            & & (0.375) & (0.339) & (0.305) & (0.294) \\ \hline
        \end{tabular}
        \caption{
        Simulation results.
        The means and standard errors are computed from 10000 random simulation samples.
        }
        \label{tab:simulation2}
    \end{table}

The results of simulation are presented in Tables \ref{tab:simulation1} and \ref{tab:simulation2}.
In each setting of simulation, we use the sample with $N = 2000$, 4000, 8000, and 10000, and standard errors are calculated from 10000
replications. 
Prior to each estimation, we have confirmed that Assumptions \ref{asm: ukul} and \ref{asm: rightinverse} are satisfied so that the matrices we use are full rank.
We use nine patterns of initial values of the parameters as follows: initial values of $\alpha_{0}$ and $\alpha_{1}$ are 0.95 times of its each true value and initial values of $\delta$ and $\beta$ are drawn from the set $\left\{ 0.7, 0.8, 0.9 \right\}$.   

Our results in Tables \ref{tab:simulation1} and \ref{tab:simulation2} suggest that our proposed estimation method does a good job across different set of sample sizes.
There are several things to notice.
First, the parameters of instantaneous utility functions are precisely estimated with small standard errors.
The estimates are stable across different sample sizes.
Second, the estimates of discount functions approach the true values as the sample size increases.
The standard errors of the discount functions are relatively larger than those of instantaneous utility functions.

\section{Conclusion} \label{sec:conclusion}

We study identification of dynamic discrete choice models with hyperbolic discounting.
We show that the standard discount factor, present bias factor, and instantaneous utility functions for the sophisticated agent are point-identified from observed CCPs and transition probabilities in a finite horizon model. 
The main idea to achieve identification is to exploit variation of the observed CCPs over time.
We also propose the estimation method and demonstrate a good performance of our estimator by simulation.

\singlespacing
\setlength\bibsep{0pt}
\bibliographystyle{econ-jpe}
\bibliography{main}

@article{abbring2010identification,
  title={Identification of dynamic discrete choice models},
  author={Abbring, Jaap H},
  journal={Annual Review of Economics},
  volume={2},
  number={1},
  pages={367--394},
  year={2010},
  publisher={Annual Reviews}
}

@techreport{abbring2018identifying,
  title={Identifying present-biased discount functions in dynamic discrete choice models},
  author={Abbring, Jaap H and Daljord, {\O}ystein and Iskhakov, Fedor},
  year={2019},
  institution={Working paper, Tilburg University. [472]}
}

@article{abbring2020comment,
  title={A comment on “Estimating dynamic discrete choice models with hyperbolic discounting” by Hanming Fang and Yang Wang},
  author={Abbring, Jaap H and Daljord, {\O}ystein},
  journal={International Economic Review},
  volume={61},
  number={2},
  pages={565--571},
  year={2020},
  publisher={Wiley Online Library}
}

@article{abbring2020identifying,
  title={Identifying the discount factor in dynamic discrete choice models},
  author={Abbring, Jaap H and Daljord, {\O}ystein},
  journal={Quantitative Economics},
  volume={11},
  number={2},
  pages={471--501},
  year={2020}
}

@article{aguirregabiria2010dynamic,
  title={Dynamic discrete choice structural models: A survey},
  author={Aguirregabiria, Victor and Mira, Pedro},
  journal={Journal of Econometrics},
  volume={156},
  number={1},
  pages={38--67},
  year={2010},
  publisher={Elsevier}
}

@article{aguirregabiria2020identification,
  title={Identification and estimation of dynamic games when players’ beliefs are not in equilibrium},
  author={Aguirregabiria, Victor and Magesan, Arvind},
  journal={Review of Economic Studies},
  volume={87},
  number={2},
  pages={582--625},
  year={2020},
  publisher={Oxford University Press}
}

@article{an2021dynamic,
  title={Dynamic decisions under subjective expectations: A structural analysis},
  author={An, Yonghong and Hu, Yingyao and Xiao, Ruli},
  journal={Journal of Econometrics},
  volume={222},
  number={1},
  pages={645--675},
  year={2021},
  publisher={North-Holland}
}

@article{arcidiacono2011conditional,
  title={Conditional choice probability estimation of dynamic discrete choice models with unobserved heterogeneity},
  author={Arcidiacono, Peter and Miller, Robert A},
  journal={Econometrica},
  volume={79},
  number={6},
  pages={1823--1867},
  year={2011},
  publisher={Wiley Online Library}
}

@article{arcidiacono2020identifying,
  title={Identifying dynamic discrete choice models off short panels},
  author={Arcidiacono, Peter and Miller, Robert A},
  journal={Journal of Econometrics},
  volume={215},
  number={2},
  pages={473--485},
  year={2020},
  publisher={North-Holland}
}

@techreport{bajari2015identification,
  title={Identification and efficient semiparametric estimation of a dynamic discrete game},
  author={Bajari, Patrick and Chernozhukov, Victor and Hong, Han and Nekipelov, Denis},
  year={2015},
  institution={National Bureau of Economic Research}
}

@article{bayer2016dynamic,
  title={A dynamic model of demand for houses and neighborhoods},
  author={Bayer, Patrick and McMillan, Robert and Murphy, Alvin and Timmins, Christopher},
  journal={Econometrica},
  volume={84},
  number={3},
  pages={893--942},
  year={2016},
  publisher={Wiley Online Library}
}

@article{de2019subsidies,
  title={Subsidies and time discounting in new technology adoption: Evidence from solar photovoltaic systems},
  author={De Groote, Olivier and Verboven, Frank},
  journal={American Economic Review},
  volume={109},
  number={6},
  pages={2137--72},
  year={2019}
}

@article{fang2009time,
  title={Time-inconsistency and welfare program participation: Evidence from the NLSY},
  author={Fang, Hanming and Silverman, Dan},
  journal={International Economic Review},
  volume={50},
  number={4},
  pages={1043--1077},
  year={2009},
  publisher={Wiley Online Library}
}

@article{fang2015estimating,
  title={Estimating dynamic discrete choice models with hyperbolic discounting, with an application to mammography decisions},
  author={Fang, Hanming and Wang, Yang},
  journal={International Economic Review},
  volume={56},
  number={2},
  pages={565--596},
  year={2015},
  publisher={Wiley Online Library}
}

@article{frederick2002time,
  title={Time discounting and time preference: A critical review},
  author={Frederick, Shane and Loewenstein, George and O'Donoghue, Ted},
  journal={Journal of Economic Literature},
  volume={40},
  number={2},
  pages={351--401},
  year={2002}
}

@article{hotz1993conditional,
  title={Conditional choice probabilities and the estimation of dynamic models},
  author={Hotz, V Joseph and Miller, Robert A},
  journal={Review of Economic Studies},
  volume={60},
  number={3},
  pages={497--529},
  year={1993},
  publisher={Wiley-Blackwell}
}

@article{hu2012nonparametric,
  title={Nonparametric identification of dynamic models with unobserved state variables},
  author={Hu, Yingyao and Shum, Matthew},
  journal={Journal of Econometrics},
  volume={171},
  number={1},
  pages={32--44},
  year={2012},
  publisher={Elsevier}
}

@article{kasahara2009nonparametric,
  title={Nonparametric identification of finite mixture models of dynamic discrete choices},
  author={Kasahara, Hiroyuki and Shimotsu, Katsumi},
  journal={Econometrica},
  volume={77},
  number={1},
  pages={135--175},
  year={2009},
  publisher={Wiley Online Library}
}

@article{keane1997career,
  title={The career decisions of young men},
  author={Keane, Michael P and Wolpin, Kenneth I},
  journal={Journal of Political Economy},
  volume={105},
  number={3},
  pages={473--522},
  year={1997},
  publisher={The University of Chicago Press}
}

@article{kennan2011effect,
  title={The effect of expected income on individual migration decisions},
  author={Kennan, John and Walker, James R},
  journal={Econometrica},
  volume={79},
  number={1},
  pages={211--251},
  year={2011},
  publisher={Wiley Online Library}
}

@article{laibson1997golden,
  title={Golden eggs and hyperbolic discounting},
  author={Laibson, David},
  journal={Quarterly Journal of Economics},
  volume={112},
  number={2},
  pages={443--478},
  year={1997},
  publisher={MIT Press}
}

@article{magnac2002identifying,
  title={Identifying dynamic discrete decision processes},
  author={Magnac, Thierry and Thesmar, David},
  journal={Econometrica},
  volume={70},
  number={2},
  pages={801--816},
  year={2002},
  publisher={JSTOR}
}

@article{norets2014semiparametric,
  title={Semiparametric inference in dynamic binary choice models},
  author={Norets, Andriy and Tang, Xun},
  journal={Review of Economic Studies},
  volume={81},
  number={3},
  pages={1229--1262},
  year={2014},
  publisher={Oxford University Press}
}

@article{o1999doing,
  title={Doing it now or later},
  author={O'Donoghue, Ted and Rabin, Matthew},
  journal={American Economic Review},
  volume={89},
  number={1},
  pages={103--124},
  year={1999}
}

@article{o2001choice,
  title={Choice and procrastination},
  author={O'Donoghue, Ted and Rabin, Matthew},
  journal={Quarterly Journal of Economics},
  volume={116},
  number={1},
  pages={121--160},
  year={2001},
  publisher={MIT Press}
}

@article{paserman2008job,
  title={Job search and hyperbolic discounting: Structural estimation and policy evaluation},
  author={Paserman, M Daniele},
  journal={The Economic Journal},
  volume={118},
  number={531},
  pages={1418--1452},
  year={2008},
  publisher={Oxford University Press Oxford, UK}
}

@article{phelps1968second,
  title={On second-best national saving and game-equilibrium growth},
  author={Phelps, Edmund S and Pollak, Robert A},
  journal={Review of Economic Studies},
  volume={35},
  number={2},
  pages={185--199},
  year={1968},
  publisher={JSTOR}
}

@article{rust1987optimal,
  title={Optimal replacement of GMC bus engines: An empirical model of Harold Zurcher},
  author={Rust, John},
  journal={Econometrica},
  pages={999--1033},
  year={1987},
  publisher={JSTOR}
}

@article{rust1994structural,
  title={Structural estimation of Markov decision processes},
  author={Rust, John},
  journal={Handbook of Econometrics},
  volume={4},
  pages={3081--3143},
  year={1994},
  publisher={Elsevier}
}

@article{strotz1955myopia,
  title={Myopia and inconsistency in dynamic utility maximization},
  author={Strotz, Robert Henry},
  journal={Review of Economic Studies},
  volume={23},
  number={3},
  pages={165--180},
  year={1955},
  publisher={JSTOR}
}

@article{thaler1981some,
  title={Some empirical evidence on dynamic inconsistency},
  author={Thaler, Richard},
  journal={Economics Letters},
  volume={8},
  number={3},
  pages={201--207},
  year={1981},
  publisher={Elsevier}
}

@article{wang2022identification,
  title={Identification of Dynamic Discrete Choice Models with Hyperbolic Discounting Using a Terminating Action},
  author={Wang, Chao and Weiergraeber, Stefan and Xiao, Ruli},
  year={2022},
  publisher={CAEPR WORKING PAPER SERIES (2022-010)}
}

\clearpage

\onehalfspacing


\clearpage



\clearpage

\section*{Appendix A. Identification with Additional State Variables} \label{sec:additional}
\addcontentsline{toc}{section}{Appendix}

In some application, the state space can be large. 
Then, by Theorem \ref{thm: onestate-identification-betadelta}, the discount functions may not be identified because the required number of time periods in the data is too large.
In this section, we introduce an additional state variable that serves to weaken this requirement.

Let us introduce another state variable $w \in \wcal \coloneqq \left\{ 1,2,\ldots,M \right\}$ ($M<\infty$).
This state variable evolves under the transition rule $h \left( w' \mid w \right)$, independently of the agent's choice.
One example for such a variable is a macro-level variable as introduced in \citet*{an2021dynamic}.
Since $w$ is a macro-level variable such as GDP or the stock price, an individual or a firm cannot affect its transition.
Note that the agent's instantaneous utility function now depends on the state variables $x$ and $w$, so the value functions, the observed CCPs, and the perceived CCPs also depend on $x$ and $w$. 

If we follow the same derivation as equation \eqref{eq: key-scalar-valuefunction}, we have
\begin{align}
    V_{t} \xwb
    = 
    &-\log \left[ P_{t,K} \xwb \right] + u_{K} \xwb + \beta \delta \sumx \sumw V_{t+1} \left( x', w' \right) f \left( x' \mid x, K \right) h \left( w' \mid w \right)
    \nonumber \\
    &+ \left( 1-\beta \right) \delta \sumj P_{t,j} \xwb \left[ \sumx \sumw V_{t+1} \left( x', w' \right) f \left( x' \mid x, j \right) h \left( w' \mid w \right) \right].
    \label{eq: key-scalar-valuefunction-additional}
\end{align}

To obtain the identification result, we need to impose the following assumption on the observed CCP:
\begin{asm}
    The observed CCP does not depend on the state variable $w$, that is,
    \begin{align}
        P_{t,i} \xwb
        = P_{t,i} \xb,
        \nonumber
    \end{align}
    for all $x,w,i$, and $t = 1,\ldots,T$.
    \label{asm: perceivedCCP-timeinvariant-additional}
\end{asm} 

This assumption can be viewed as follows: the additional state variable $w$ does not affect the choice probability.
For example, let us consider the labor force participation problem and let $w$ be a macro-level variable such as a nation's GDP.
The agent does not take the macro-level variable $w$ into consideration to decide labor force participation.
Note that the instantaneous utility function may depend on both $x$ and $w$.
This does not contradict this assumption as long as the proportion $\frac{\exp \left[ W_{t,i} \xwb \right]}{\sum_{j=1}^{K} \exp \left[ W_{t,j} \xwb \right]}$, which is equal to the observed CCP with an additional state variable, is the same as the proportion $\frac{\exp \left[ W_{t,i} \xb \right]}{\sum_{j=1}^{K} \exp \left[ W_{t,j} \xb \right]}$, which is equal to the observed CCP without the additional state variable.

Under Assumption \ref{asm: perceivedCCP-timeinvariant-additional}, we have the following equation:
\begin{align}
    \vv_{t}
    =
    - \log \left( \pp_{t,K} \right) + \mathbf{u}_{K} + \bd \tilde{\ff}_{K} \vt \hf + \left( 1-\beta \right) \delta \pp_{t} \ff \vt \hf
    - \left( 1-\beta \right) \delta \pp_{t,J} \ff_{J} \vt \hf,
    \nonumber
\end{align}
where $\hf \wb = \left[ h \left( w'=1 \mid w \right), \ldots, h \left( w'=M \mid w \right) \right]'$ is an $M \times 1$ vector, $\hf = \left[ \hf \left( w=1 \right), \ldots, \hf \left( w=M \right) \right]$ is an $M \times M$ matrix, $\vv_{t}$ is now an $(J-1) \times M$ matrix of the form:
\begin{align}
    \vv_{t}
    =
    \left[
    \begin{array}{ccc}
        V_{t} \left( x'=1, w'=1 \right) & \ldots & V_{t} \left( x'=1, w'=M \right) \\
        \vdots & \ddots & \vdots \\
        V_{t} \left( x'=J-1, w=1 \right) & \ldots & V_{t} \left( x'=J-1, w=M \right)
    \end{array}
    \right],
    \nonumber
\end{align}
and the other matrices and vectors, such as $\log \left( \pp_{t,K} \right)$, $\mathbf{u}_{K}$, $\tilde{\ff}_{K}$, $\pp_{t}$, $\ff$, $\pp_{t,J}$, and $\ff_{J}$, are the same as in Section \ref{sec: identification}.
Note that, to obtain this matrix form, Assumption \ref{asm: perceivedCCP-timeinvariant-additional} is crucial.
Otherwise, the product of the observed CCP matrix $\pp_{t}$ and the transition matrix $\ff$ cannot be well-defined (the product of $\pp_{t,J}$ and $\ff_{J}$ cannot be well-defined, either).

By taking the first difference and multiplying the matrix $\hf$ from the right, we obtain
\begin{align}
    \de \vv_{t} \hf
    =
    &- \de \log \left( \pp_{t,K} \right) \hf + \bd \tilde{\ff}_{K} \de \vt \hf \hf 
    \nonumber \\
    &+ \left( 1-\beta \right) \delta
    \left[ \left( \pp_{t} \ff - \pp_{t,J} \ff_{J}  \right) \vt \hf \hf - \left( \pp_{t-1} \ff - \pp_{t-1,J} \ff_{J} \right) \vv_{t} \hf \hf \right].
    \label{eq: key-vector-difference-valuefunction-additional}
\end{align}

As in Section \ref{sec: identification}, we consider the CCP ratio:
\begin{align}
    &\ 
    D_{t,k,\ell} \left( x_{1}, x_{2}, w \right)
    \nonumber \\
    &\coloneqq \log \left( \frac{P_{t,k} \left( x_{1}, w \right)}{P_{t,\ell} \left( x_{2}, w \right)} \right)
    \nonumber \\
    &= W_{t,k} \left( x_{1}, w \right) - W_{t,\ell} \left( x_{2}, w \right)
    \nonumber \\
    &= \left[ u_{k} \left( x_{1}, w \right) - u_{\ell} \left( x_{2}, w \right) \right] + \bd \sumx \sumw V_{t+1} \left( x',w' \right) \left[ f \left( x' \mid x_{1}, k \right) - f \left( x' \mid x_{2}, \ell \right) \right] h \left( w' \mid w \right)
    \nonumber \\
    &= \left[ u_{k} \left( x_{1}, w \right) - u_{\ell} \left( x_{2}, w \right) \right] + \bd \left[ \ff_{k} \left( x_{1} \right) - \ff_{\ell} \left( x_{2} \right) \right] \vt \hf \wb,
    \label{eq: CCPratio-additional}
\end{align}
where $\ff_{i} \xb$ is the same $1 \times (J-1)$ vector as in Section \ref{sec: identification}.

To obtain further transformations, we need the following assumption:
\begin{asm}
    (a) There exist at least $J-1$ action pairs $\left( k^{j}, \ell^{j} \right)$ and state pairs $\left( x_{1}^{j}, x_{2}^{j} \right)$ such that: 
    \begin{align}
        u_{k^{j}} \left( x_{1}^{j}, w \right) = u_{\ell^{j}} \left( x_{2}^{j}, w \right)
        \ 
        \text{ for all } w \in \wcal,
    \end{align}
    with either (i) $k^{j} \ne \ell^{j}$, (ii) $x_{1}^{j} \ne x_{2}^{j}$, or (iii) both, for $j = 1, \ldots, J-1$.
    \\
    (b) the matrix $\ftil$ is full column rank.
    \label{asm: ukul-add}
\end{asm}

Note that the matrix $\ftil$ is the same $\left( J-1 \right) \times \left( J-1 \right)$ matrix as in Section \ref{sec: identification} and that this assumption corresponds to Assumption \ref{asm: ukul}.
The only difference between them is that, in Assumption \ref{asm: ukul-add}, the instantaneous utility depends not only on the state variable $x$ but also on the state variable $w$.

By Assumption \ref{asm: ukul-add}, we can have the following expressions:
\begin{align}
    \vt \hf
    &= \bdinv \ftil^{-1} \dd_{t}
    \label{eq: key2-V-afterinv-additional}
    \\
    \de \vt \hf
    &= \bdinv \ftil^{-1} \de \dd_{t},
    \label{eq: key2-deltaV-afterinv-additional}
\end{align}
where $\dd_{t} = \left[ D_{t,k^{1},\ell^{1}} \left( x_{1}^{1}, x_{2}^{1} \right), \ldots, D_{t,k^{J-1},\ell^{J-1}} \left( x_{1}^{J-1}, x_{2}^{J-1} \right) \right]'$ is a $(J-1) \times 1$ vector, as in Section \ref{sec: identification}.
By substituting \eqref{eq: key2-deltaV-afterinv-additional} and \eqref{eq: key2-V-afterinv-additional} into \eqref{eq: key-vector-difference-valuefunction-additional}, we obtain
\begin{align}
    &\ 
    \bdinv \ftil^{-1} \de \dd_{t-1}
    \nonumber \\
    &=
    - \de \log \left( \pp_{t,K} \right) \hf 
    + \tilde{\ff}_{K} \ftil^{-1} \de \dd_{t} \hf + \bfrac \left( \pp_{t} \ff - \pp_{t,J} \ff_{J} \right) \ftil^{-1} \dd_{t} \hf 
    \nonumber \\
    &\ - \bfrac \left( \pp_{t-1} \ff - \pp_{t-1,J} \ff_{J} \right) \ftil^{-1} \dd_{t-1} \hf,
    \label{eq: key3-longform-additional}
\end{align}
for $t=3,\ldots,T$.
This relationship can be equivalently expressed as
\begin{align}
    \left[
    \begin{array}{ccc}
        \ii_{J-1} & \bfrac \ii_{J-1} & - \bdinv \ii_{J-1}
    \end{array}
    \right]
    \abt
    =
    \Bt,
    \label{eq: key3-matrixform-additional}
\end{align}
where 
\begin{align}
    \abt
    &=
    \left[ 
    \begin{array}{ccc}
        \tilde{\ff}_{K} \ftil^{-1} \de \dd_{3} \hf & \cdots & \tilde{\ff}_{K} \ftil^{-1} \de \dd_{T} \hf \\
        \pf_{3} \hf & \cdots & \pf_{T} \hf \\
        \ftil^{-1} \de \dd_{2} & \cdots & \ftil^{-1} \de \dd_{T-1}
    \end{array}
    \right]
    \nonumber
\end{align}
is a $3(J-1) \times (T-2)M$ matrix where 
\begin{align}
    \pf_{t} = \left( \pp_{t} \ff - \pp_{t,J} \ff_{J}  \right) \ftil^{-1} \dd_{t} - \left( \pp_{t-1} \ff - \pp_{t-1,J} \ff_{J} \right) \ftil^{-1} \dd_{t-1}
    \nonumber
\end{align}
and
$\Bt = \left[ \de \log \left( \pp_{3,K} \right) \hf, \ldots, \de \log \left( \pp_{T,K} \right) \hf \right]$ is a $(J-1) \times (T-2)M$ matrix.

For the identification result, we introduce the following assumption:
\begin{asm}
    (a) The number of observed periods $T$ satisfies $(T-2)M \ge 3(J-1)$.
    \\
    (b) The matrix $\abt$ is full row rank. 
    \label{asm: rightinverse-additional}
\end{asm}

This assumption corresponds to Assumption \ref{asm: rightinverse} in Section \ref{sec: identification}.
Under Assumption \ref{asm: rightinverse-additional}, the matrix $\abt$ has the right inverse matrix. 
Thus, both $\bfrac \ii_{J-1}$ and $-\bdinv \ii_{J-1}$ in equation \eqref{eq: key3-matrixform-additional} have a closed-form solution.
This assumption is empirically testable.

By the similar logic to Theorem \ref{thm: onestate-identification-betadelta}, we have the following theorem for the model with the additional state variables:

\begin{thm}
    Suppose that Assumptions \ref{asm: transition}, \ref{asm: logit-error}, \ref{asm: additive-separability}, \ref{asm: perceivedCCP-timeinvariant-additional}, \ref{asm: ukul-add}, and \ref{asm: rightinverse-additional} hold.
    Then the discount functions $\beta$ and $\delta$ and the perceived CCP matrix $\ptil_{t}$ are identified for $t=1,2,\ldots,T$ with $T \ge \lceil \frac{3(J-1)}{M} \rceil + 2$, where $\lceil x \rceil$ is the minimum integer larger than or equal to $x$.
    Furthermore, if the utility level of the reference action for each state is known, then the instantaneous utility functions $u_{i} \xwb$ for all $i \in \ical$, all $x \in \xcal$, and all $w \in \wcal$ are identified.
    \label{thm: additional-identification-betadelta}
\end{thm}
The proof is similar to Theorem \ref{thm: onestate-identification-betadelta}, so we omit it.
According to Theorem \ref{thm: additional-identification-betadelta}, the present-bias factor $\beta$ and the standard discount factor $\delta$ are identified in the form of functions of the observed CCPs and the transition density.
Also, due to the stationary structure of the instantaneous utility, each instantaneous utility function is identified. 
If we compare it with Theorem \ref{thm: onestate-identification-betadelta}, we find that the required data length is reduced if we have an additional state variable in the data.
For example, in the DDC model with the (original) state variable takes on three values ($J=3$), if there is an additional state variable that takes on two values ($M=3$), we only need four periods of data.
On the other hand, without such an additional state variable, we would need eight periods of data (see Theorem \ref{thm: onestate-identification-betadelta}).

\section*{Appendix B. Proofs} \label{sec: proofs}
\addcontentsline{toc}{section}{Appendix}

\subsection*{Derivation of equation \eqref{eq: CCPratio}}

\begin{align}
    \sumx V_{t+1} \xbn f \left( x' \mid x, i \right)
    &= \sumxm V_{t+1} \xbn f \left( x' \mid x, i \right) + V_{t+1} \jb f \left( J \mid x, i \right)
    \nonumber \\
    &= \sumxm V_{t+1} \xbn f \left( x' \mid x, i \right) + V_{t+1} \jb \left[ 1 - \sumxm f \left( x' \mid x, i \right) \right]
    \nonumber \\
    &= \sumxm \left[ V_{t+1} \xbn - V_{t+1} \jb \right] f \left( x' \mid x, i \right) + V_{t+1} \jb
    \nonumber \\
    &= \ff_{i} \xb \vt + V_{t+1} \jb,
    \nonumber
\end{align}
where $\ff_{i} \xb$ is a $1 \times (J-1)$ vector and $\vt$ is a $(J-1) \times 1$ vector.

\subsection*{Proof of Theorem \ref{thm: onestate-identification-betadelta}}
Using Assumption \ref{asm: rightinverse}, we have that $\ab$ has a right inverse matrix $\ab^{+}$.
Thus we have
\begin{align}
    \left[
    \begin{array}{ccc}
        \ii_{J-1} & \bfrac \ii_{J-1} & - \bdinv \ii_{J-1}
    \end{array}
    \right]
    = \left[ \de \log \left( \pp_{K} \right) \right] \ab^{+}.
    \nonumber
\end{align}
Note that $\bfrac \ii_{J-1}$ and $-\bdinv \ii_{J-1}$ are identified separately because we have an explicit form of the right-hand side matrix.
Since each element of $\bfrac \ii_{J-1}$ is known, the fraction $\bfrac$ is identified and so is $\beta$.
Since $-\bdinv \ii_{J-1}$ has already been identified and thus $\bd$ is also identified, we can identify $\delta$.


Next, let us recall that the instantaneous utility function is time-invariant and focus on the terminal period.
Suppose that, without loss of generality, the value of $u_{K} \left(J\right)$ is known.
By equation \eqref{eq: CCPratio}, we have, for any choice $i \in \ical$ and any state $x \in \xcal$,
\begin{align}
    W_{T,i} \xb - W_{T,K} \left(J\right)
    = u_{i} \xb - u_{K} \left(J\right),
    \label{eq: identify-u-terminal}
\end{align}
because the continuation value at the terminal period is set to be zero, as described in Section \ref{sec:estimation}.
Since the value of $u_{K} \left(J\right)$ is known, $u_{i} \xb$ is uniquely determined because the left-hand side of equation \eqref{eq: identify-u-terminal} is the log of the CCP ratio and thus known.
This completes the proof.

\subsection*{Derivation of equation \eqref{eq: CCPratio-additional}}

\begin{align}
    &\ 
    \sumx \sumw V_{t+1} \left( x', w' \right) f \left( x' \mid x, i \right) h \left( w' \mid w \right)
    \nonumber \\
    &=
    \sum_{x'=1}^{J-1} \sumw V_{t+1} \left( x', w' \right) f \left( x' \mid x, i \right) h \left( w' \mid w \right)
    + \sumw V_{t+1} \left( J, w' \right) f \left( J \mid x,i \right) h \left( w' \mid w \right)
    \nonumber \\
    &=
    \sum_{x'=1}^{J-1} \sumw V_{t+1} \left( x', w' \right) f \left( x' \mid x, i \right) h \left( w' \mid w \right)
    + \sumw V_{t+1} \left( J, w' \right) \left[ 1 - \sum_{x'=1}^{J-1} f \left( x' \mid x,i \right) \right] h \left( w' \mid w \right)
    \nonumber \\
    &= \sum_{x'=1}^{J-1} \sumw \left[ V_{t+1} \left( x', w' \right) - V_{t+1} \left( J, w' \right) \right] f \left( x' \mid x, i \right) h \left( w' \mid w \right) 
    + \sumw V_{t+1} \left( J, w' \right) h \left( w' \mid w \right)
    \nonumber \\
    &=
    \ff_{i} \xb \vt \hf \wb 
    + \sumw V_{t+1} \left( J, w' \right) h \left( w' \mid w \right),
    \nonumber
\end{align}
where $\ff_{i} \xb$ is a $1 \times (J-1)$ vector, $\vt$ is a $(J-1) \times M$ matrix, and $\hf \wb$ is an $M \times 1$ vector.

\end{document}